\documentclass[twocolumn,showpacs,preprintnumbers,amsmath,amssymb]{revtex4}
\usepackage{epsfig,amsfonts}
\usepackage{amsmath}
\usepackage{bbm}
\usepackage{graphicx,psfrag,rotating}
\usepackage{graphics}
\usepackage{txfonts}

\usepackage{graphicx}
\usepackage{dcolumn}
\usepackage{bm}

\begin{document}


\title{Creating and detecting micro-macro photon-number entanglement by amplifying and de-amplifying a single-photon entangled state}

\author{Roohollah Ghobadi$^{1,2}$, Alexander Lvovsky$^1$, and  Christoph Simon$^1$}
\affiliation{$^1$ Institute for Quantum Information Science
and Department of Physics and Astronomy, University of
Calgary, Calgary T2N 1N4, Alberta, Canada\\$^2$Department of Physics, Sharif University of Technology, Tehran, Iran}

\date{\today}

\begin{abstract}
We propose a scheme for the observation of micro-macro entanglement in photon number based on amplifying and de-amplifying a single-photon entangled state in combination with homodyne quantum state tomography. The created micro-macro entangled state, which exists between the amplification and de-amplification steps, is a superposition of two components with mean photon numbers that differ by approximately a factor of three. We show that for reasonable values of photon loss it should be possible to detect micro-macro photon-number entanglement where the macro system has a mean number of one hundred photons or more.
\end{abstract}

\maketitle

The goal of pushing the observation of quantum effects such as superpositions and entanglement towards the macroscopic level is currently being pursued in a number of different fields, including  trapped ions \cite{Monroe96}, superconducting circuits \cite{Friedman00}, nano-mechanics \cite{Connell10}, microwave cavities interacting with atoms in Rydberg states \cite{Deleglise08}, atomic ensembles \cite{polzik}, and non-linear optics \cite{yurke,entlaser,Ourjoumtsev07,Demartini08}. Within non-linear optics, one can distinguish proposals based on Kerr non-linearities \cite{yurke}, and proposals and experiments based on parametric down-conversion \cite{entlaser,Ourjoumtsev07,Demartini08}. The latter area has recently seen significant activity, a lot of which was stimulated by Ref. \cite{Demartini08}, which claimed the creation and detection of entanglement in polarization between a single photon on one side and thousands of photons on the other. The state was created starting from a single polarization entangled photon pair, by greatly amplifying one of the photons using stimulated type-II parametric down-conversion (i.e. a two-mode squeezing interaction involving both polarization modes). Ref. \cite{Sekatski} subsequently showed that the evidence for micro-macro entanglement given in Ref. \cite{Demartini08} was not conclusive, and Ref. \cite{Raeisi11} showed that in order to rigorously prove the presence of entanglement for the state of Ref. \cite{Demartini08}, one would need to be able to count the photons on the macro side with single-photon resolution, which is a significant technological challenge. Several other results also suggest that the observation of entanglement by direct measurement on macroscopic systems generally requires very high resolution, which also implies very low photon loss in the case of multi-photon states \cite{mermin,peres,entlaser,sciarrino}.

These results inspired the work of Ref. \cite{Raeisi12}, which proposed to prove the existence of micro-macro polarization entanglement by de-amplifying the macro part of the state of Ref. \cite{Demartini08} back to the single-photon level. As entanglement cannot be created locally, if entanglement is detected at the single-photon level, this proves that micro-macro entanglement had to exist after the amplification stage. This approach has two advantages. On the one hand, the final measurement can be done by single-photon detection, which is much simpler than counting large photon numbers with single-photon resolution. On the other hand, the entanglement is primarily sensitive to loss between the amplification and de-amplification stages, which is easier to minimize in practice than the overall loss, which also includes detection inefficiency.

Most of the previous work in this area was concerned with polarization (or spin) entanglement \cite{entlaser,Demartini08,mermin,peres,sciarrino,Raeisi12}. Here we propose to apply the amplification-deamplification approach
to create and detect a different - quite striking - type of entanglement, namely micro-macro entanglement in photon number. Instead of starting from a polarization entangled photon pair, we propose to start from a single-photon entangled state, which can be created by sending a single photon onto a beam splitter \cite{banaszek,Lvovsky04,VanEnk,morin}. The presence of single-photon entanglement can be proven experimentally by homodyne tomography \cite{Lvovsky04}, and also by a combination of interference and single-photon detection \cite{Cho05}. Using single-photon entanglement as a starting point, one can create a micro-macro photon-number entangled state by amplifying one side via stimulated type-I parametric down-conversion (i.e. a single-mode squeezer). The resulting state is a superposition of two components with largely different mean photon numbers. This distinguishes our proposal from another recent proposal, where micro-macro entanglement is created by displacing (rather than squeezing) one half of a single-photon entangled state \cite{Sekatski12}. In that case the mean photon numbers of the two superposed components are very similar. Micro-macro photon number entanglement as suggested here could in principle be used to test proposals for fundamental decoherence in energy \cite{energy-decoherence}.

\begin{figure}
\epsfig{file=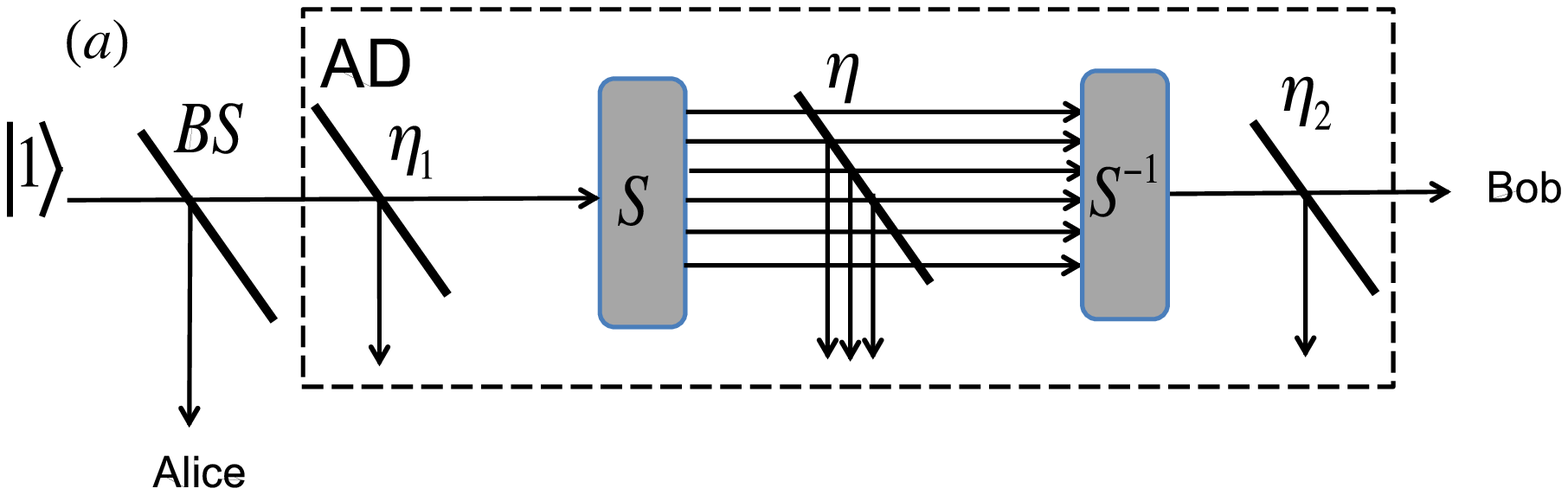,width=1\columnwidth}
\epsfig{file=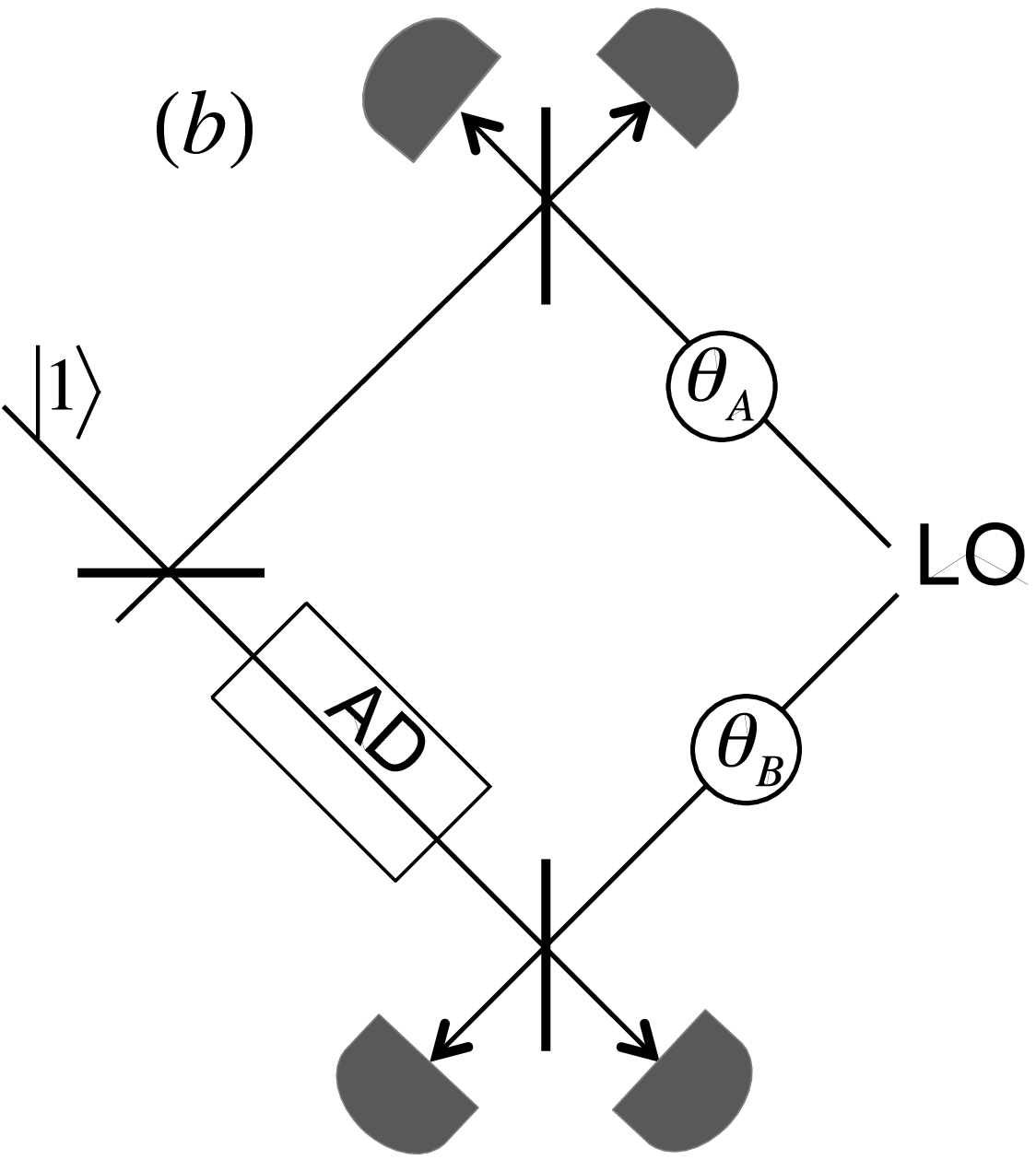,width=0.5\columnwidth}

\caption{Schematic of the proposed experiment. (a) A single photon is sent onto the beam splitter $BS$, which creates the single-photon entangled state of Eq. (1).  Mode $A$ is directly measured, see below. Mode $B$ is first amplified by the single-mode squeezer $S$ and then de-amplified by $S^{-1}$. The pump laser beams necessary for implementing $S$ and $S^{-1}$ are not shown for simplicity. Losses before $S$, between $S$ and $S^{-1}$, and after $S^{-1}$ are taken into account through transformation factors $\eta_{1}$, $\eta$ and $\eta_{2}$, respectively. If the modes $A$ and $B$ are found to be entangled at the end, one can infer the existence of micro-macro entanglement between the applications of $S$ and $S^{-1}$.
 (b) Measurement scheme including the local oscillator $LO$ which is essential for performing homodyne tomography of the final state. The box $AD$ represents the amplification and de-amplification process described in (a).
\label{fig1}
}
\end{figure}

In the following we describe our proposal in more detail, taking into account the effects of photon loss, see also Figure 1. We start by sending a single photon onto a balanced beam splitter. The output state is a single-photon entangled state
\begin{equation}
|\psi_{in}\rangle=\frac{|1\rangle_{A}|0\rangle_{B}+|0\rangle_{A}|1\rangle_{B}}{\sqrt{2}}.
\label{in}
\end{equation}
The photon in arm B is then subjected to the unitary evolution $S=e^{-iHt}$ where
\begin{equation}
H=i\chi(a^{2}-a^{\dagger^{2}})
\label{sq}
\end{equation}
is the single-mode squeezing Hamiltonian, which can be implemented by type-I parametric down conversion. The total state after the application of $S$ becomes
\begin{equation}
|\psi_{s}\rangle=\frac{|1\rangle_{A}|S_{0}\rangle_{B}+|0\rangle_{A}|S_{1}\rangle_{B}}{\sqrt{2}},
\end{equation}
where
\begin{equation}
|S_{0}\rangle=S|0\rangle=\frac{1}{\sqrt{\cosh r}}\sum_{n=0}^{\infty}\frac{\sqrt{(2n)!}}{2^{n}n!}(-\tanh r)^{n}|2n\rangle
\end{equation}
and
\begin{equation}
|S_{1}\rangle=S|1\rangle=\frac{1}{\sqrt{(\cosh r){}^{3}}}\sum_{n=0}^{\infty}\frac{\sqrt{(2n+1)!}}{2^{n}n!}(-\textrm{tanh}r)^{n}|2n+1\rangle,
\end{equation}
with $r=\chi t$.
For large enough squeezing parameter $r$ the state $|\psi_{s}\rangle$ is a superposition of two components with largely different mean photon numbers. To see this, we calculate
\begin{equation}
n_{0}=\langle S_{0}|a^{\dagger}a|S_{0}\rangle= \sinh^{2}(r)
\label{n0}
\end{equation}
and
\begin{equation}
n_{1}=\langle S_{1}|a^{\dagger}a|S_{1}\rangle=1+3\sinh^{2}(r).
\label{n1}
\end{equation}
For large enough values of $r$ one has
$n_{1}/n_{0}\sim3$. In Figure 2 we show the photon number distributions for the states $|S_0\rangle$ and $|S_1\rangle$, as well as their Wigner functions. Note that values of $r$ much greater than those used in the figure were achieved in the experiment of Ref. \cite{Demartini08}, where thousands of photons were created on the macro side.

\begin{figure}
\epsfig{file=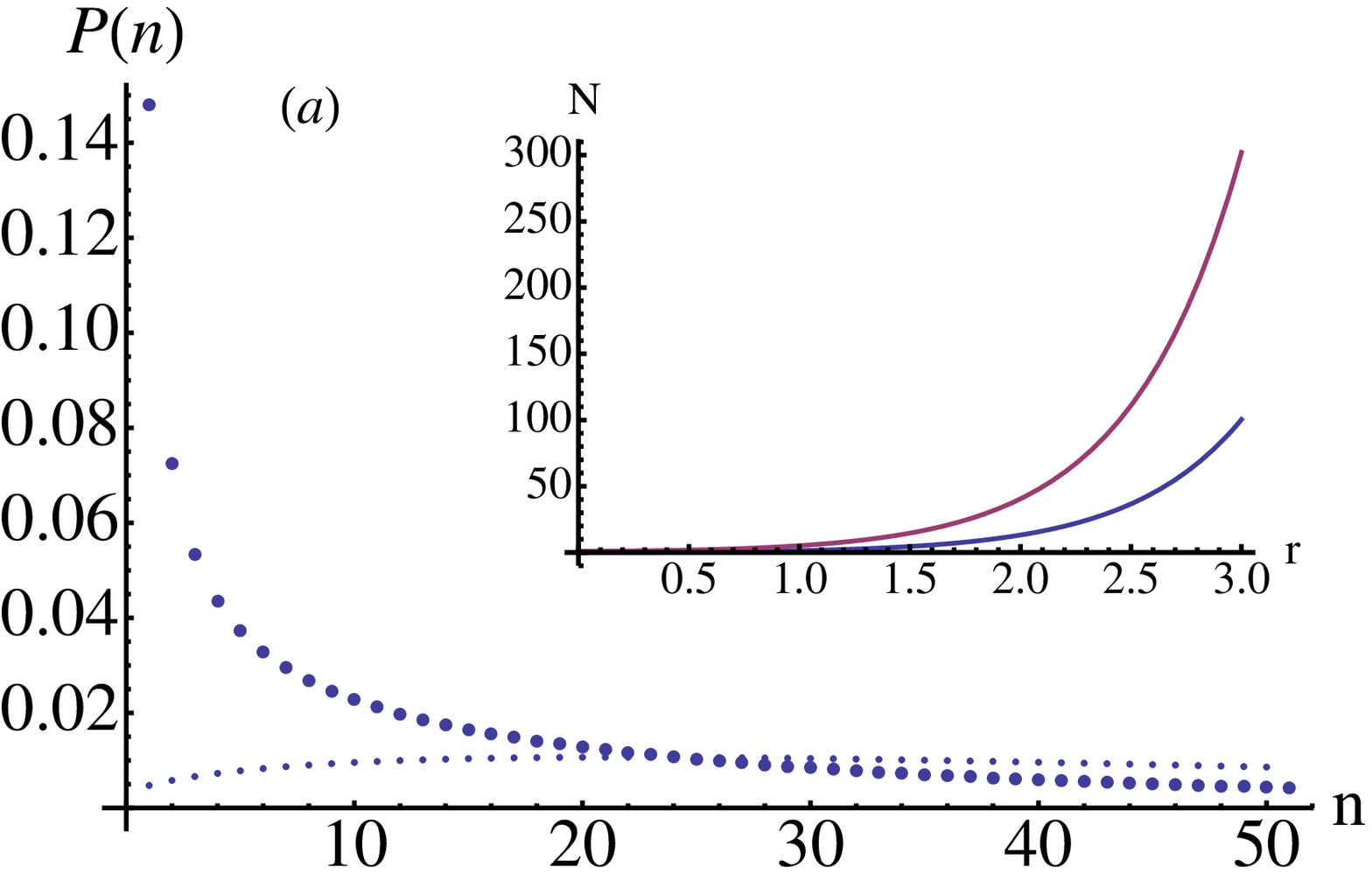,width=1\columnwidth}
\scalebox{.4}{\includegraphics*[viewport=0 0 1200 300]{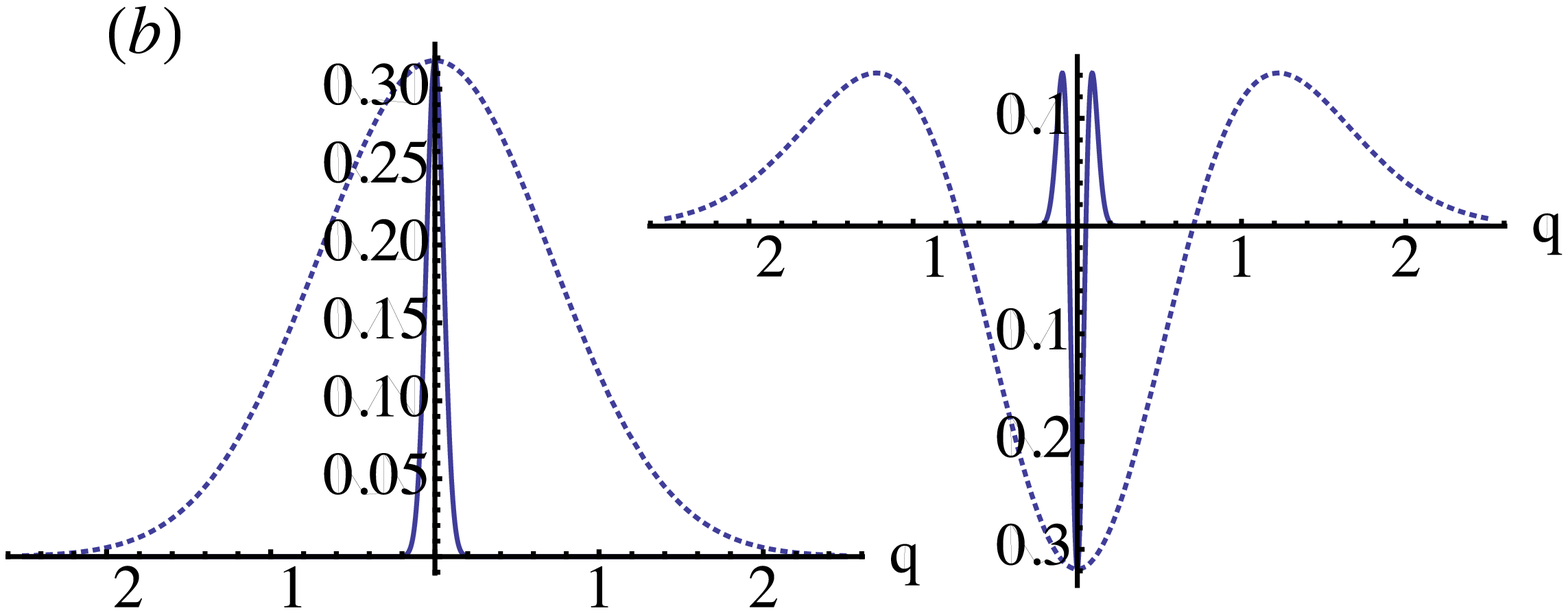}}
\caption{\label{fig2}
(a) Photon probability distributions of $|S_{0}\rangle$  (big dots) and $ |S_{1}\rangle$   (small dots) for $r=2.6$. The inset shows the mean photon numbers for the same two states as a function of $r$, see also Eqs. (\ref{n0}) and (\ref{n1}).  (b) Cross section of the Wigner function of $|S_{0}\rangle$ (left) and $|S_{1}\rangle$ (right) for the cutting plane $P=0$. The dotted curves are the Wigner functions of $|0\rangle$ and $|1\rangle $, respectively. Note that the other quadrature (which is not shown here) is correspondingly elongated due to the Heisenberg uncertainty principle.
}
\end{figure}

 So far we have discussed the amplification process. In order to study the effects of photon loss and subsequent de-amplification as shown in Figure 1, it is most convenient to work with the Wigner function. The Wigner function corresponding to the state of Eq. (\ref{in}) is given by
\begin{equation}
W_{in}(X_{A},P_{A},X_{B},P_{B})=\frac{1}{2}\sum_{m,n=0}^{m,n=1}W_{m,n}(X_{A},P_{A})W_{1-m,1-n}(X_{B},P_{B}),
\end{equation}
where
$W_{0,0}(X_{i},P_{i})=\frac{e^{-(X_{i}^{2}+P_{i}^{2})}}{\pi}$,
$W_{1,0}(X_{i},P_{i})=\frac{\sqrt{2}(X_{i}+iP_{i})e^{-(P_{i}^{2}+X_{i}^{2})}}{\pi}=W^{*}_{0,1}(X_{i},P_{i})$ and $W_{1,1}(X_{i},P_{i})=\frac{(-1+2X_{i}^{2}+2P_{i}^{2})e^{-(P_{i}^{2}+X_{i}^{2})}}{\pi}$ with $i=A,B$. $X_{i}$,$P_{i}$ are the position and momentum quadratures respectively.


The effect of squeezing in the phase space is simply given by the
following transformation: $X_{B}\rightarrow e^{r}X_{B}$, $P_{B}\rightarrow e^{-r}P_{B}$.
This implies that the Wigner function after squeezing is given by
$W_{s}(X_{A},P_{A},X_{B},P_{B})=W_{in}(X_{A},P_{A},e^{r}X_{B},e^{-r}P_{B})$.

In the absence of photon loss, the prepared macroscopic state in arm B would now undergo the de-amplification operation $S^{-1}$, which can be realized by changing the sign of $\chi$. Experimentally this can be done either by inverting the phase of the pump laser or by inverting the sign of the non-linear coefficient of the second non-linear crystal, in analogy to what is done in periodic poling \cite{Armstrong62}. In practice the de-amplification will always be preceded by a certain amount of photon loss. In this case the final state is no longer exactly equal to the initial state, in particular there will be higher order excitations in the number basis (beyond one). We will first discuss only loss between $S$ and $S^{-1}$, which is the most critical imperfection; loss before $S$ and after $S^{-1}$ will be discussed below.

The effect of loss in phase space is described by a convolution \cite{Leonhardt}
\begin{eqnarray}
&&W_{s,\eta}(X_{A},P_{A},X_{B},P_{B})= \nonumber \\  &&\intop_{-\infty}^{\infty}dX_{B}'dP_{B}'W_{s}(X_{A},P_{A},X_{B}',P_{B}')F_{\eta}(X_{B},P_{B},X_{B}',P_{B}')
\end{eqnarray}
with the  attenuation kernel
\begin{equation}
F_{\eta}(X_{B},P_{B},X_{B}',P_{B}')=\frac{\exp[-\frac{\eta}{1-\eta}((X_{B}'-\frac{X_{B}}{\sqrt{\eta}})^{2}+(P_{B}'-\frac{P_{B}}{\sqrt{\eta}})^{2}))]}{\pi(1-\eta)}.
\end{equation}
The Wigner function of the final state after de-squeezing is given
by $W_{s,\eta,s^{-1}}(X_{A},P_{A},X_{B},P_{B})=W_{s,\eta}(X_{A},P_{A},e^{-r}X_{B},e^{r}P_{B})$.
Using Eqs.[9,10] one can then obtain the density matrix of the final state\cite{Leonhardt}.

As mentioned above, on the $B$ side the final state will in general have higher-order components in the Fock basis. As it is difficult to quantify entanglement in high-dimensional systems, we will here focus on the projection of the final state onto the zero and first excitation subspace for mode $B$, i.e
\begin{equation}
\rho_{p}=(I_{A}\otimes P_{B})\rho(I_{A}\otimes P_{B}),
\end{equation}
where $I_{A}$ is the identity operator in mode $A$  and $P_{B}$ is the projection in subspace $\{|0\rangle_{B},|1\rangle_{B}\}$ in arm $B$. Since the local projection $P_B$ cannot create entanglement, any entanglement present in $\rho_p$ also had to be present in $\rho$. Similarly, any entanglement present in $\rho$ had to be present in the micro-macro state created by the amplification stage because the loss and de-amplification are also local processes.

The projected density matrix $\rho_p$ has the following form in the Fock state basis,
\begin{equation}
\rho_{p}=\left(\begin{array}{cccc}
p_{00} & 0 & 0 & d'\\
0 & p_{01} & d & 0\\
0 & d^{\ast} & p_{10} & 0\\
d'^{\ast} & 0 & 0 & p_{11}
\end{array}\right),
\label{ro10}
\end{equation}
where $p_{ij}$ is the probability to find $i$ photons in arm $A$ and $j$ photons in arm $B$; $d$ is the coherence term between $|1\rangle_{A}|0\rangle_{B}$ and $|0\rangle_{A}|1\rangle_{B}$  , $d'$ is the coherence between  $|0\rangle_{A}|0\rangle_{B}$ and $|1\rangle_{A}|1\rangle_{B}$. One should note that the projected density matrix is not normalized. In fact the success probability of projection is given by  $Tr(\rho_{p})=p_{00}+p_{01}+p_{10}+p_{11}$. In the initial state $p_{01}=p_{10}=d=\frac{1}{2}$, with all other coefficients equal to zero.
One sees that the combination of amplification, loss in between, and de-amplification can create new population terms as well as a new coherence. However, certain coherences are still exactly zero (under otherwise ideal conditions). This can be understood by noting that neither the Hamiltonian nor the loss can create coherence between neighboring photon number states (in a given mode).
 The zero elements in Eq. (\ref{ro10}) can also be understood in term of the reflection symmetry of the initial state and attenuation kernel Eq. (10).

\begin{figure}
\epsfig{file=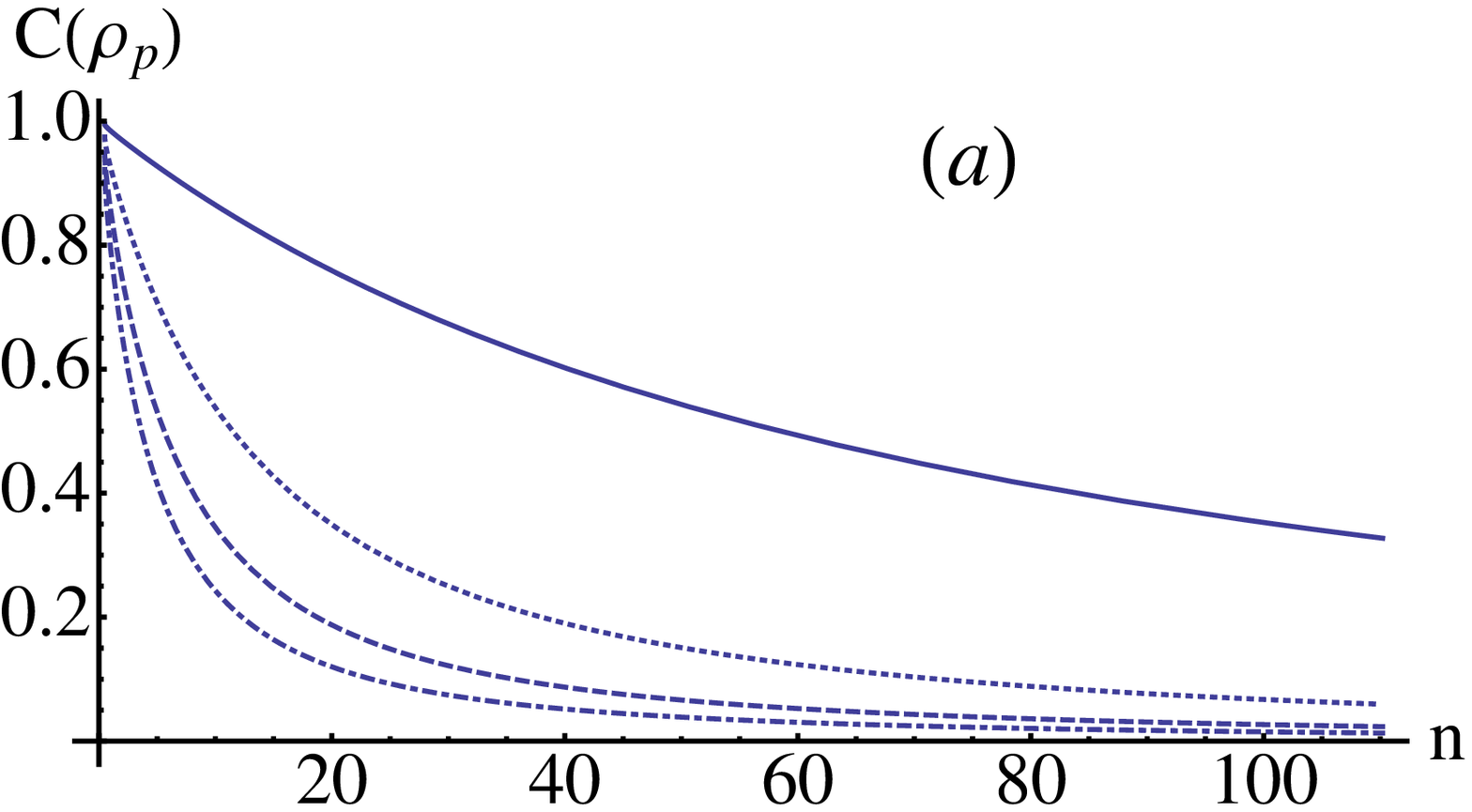,width=1\columnwidth}
\scalebox{0.5}{\includegraphics*[viewport=0 0 1200 300]{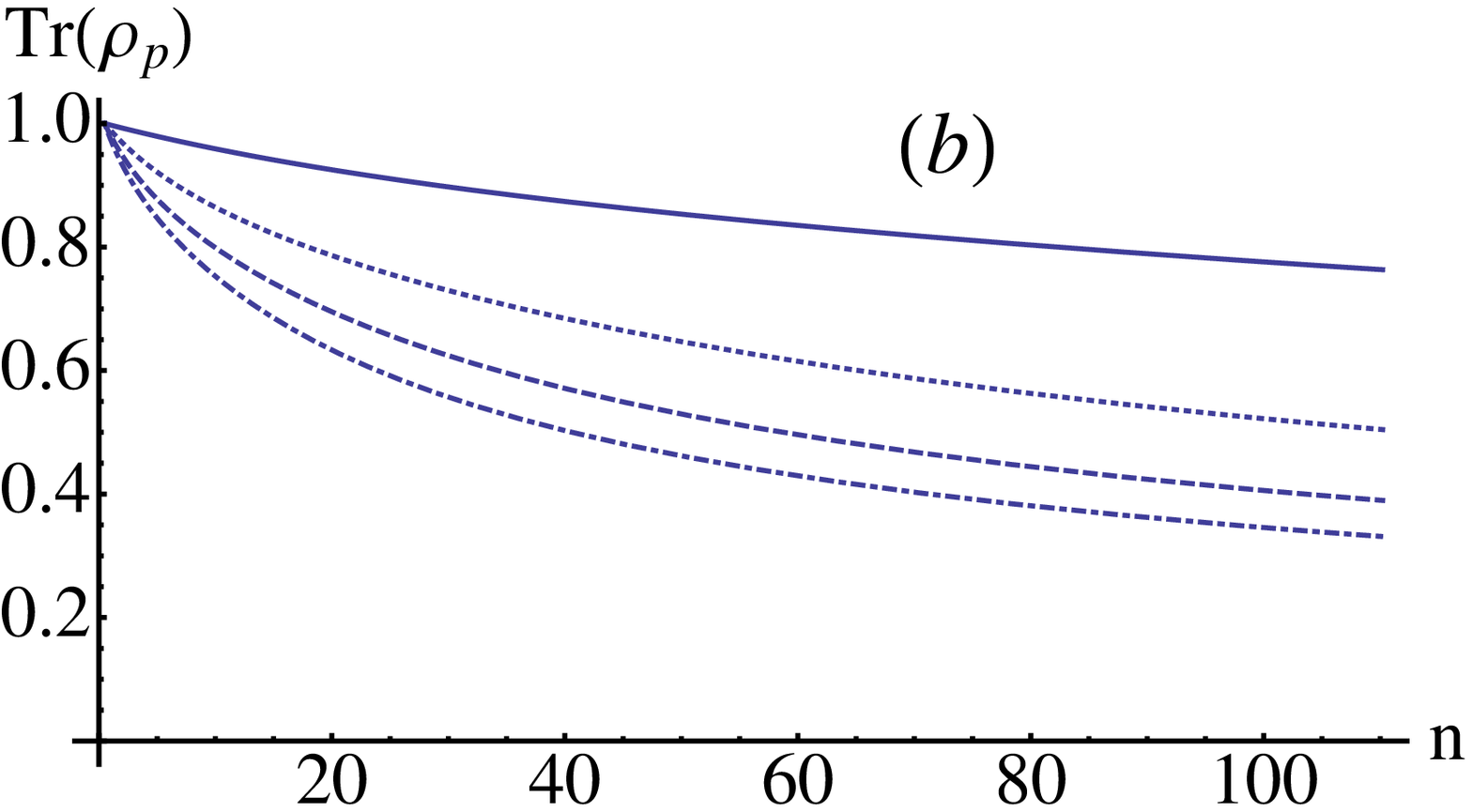}}
\caption{\label{fig3} (a) Concurrence of the final state as a function of the mean photon number after the amplification, $n=\frac{n_0+n_1}{2}$, for different values of the attenuation between amplification and de-amplification, $\eta$. The solid, dotted, dashed and dot-dashed curves correspond to $\eta=0.99, 0.95, 0.9, 0.85$, respectively.
(b) Probability of projecting the final state into the subspace spanned by the zero and one photon states for the same values of $\eta$.}
\end{figure}

To characterize entanglement we use the concurrence\cite{Wootters98} which is defined as
\begin{equation}
C(\rho_{p})=\max(0,\lambda_{1}-\lambda_{2}-\lambda_{3}-\lambda_{4})
\end{equation}
where $\lambda_{i}$ are the eigenvalues in decreasing order of the Hermitian matrix $\sqrt{\sqrt{\rho_{p}}\tilde{\rho}_{p}\sqrt{\rho_{p}}}$ with
$\tilde{\rho}_{p}=(\sigma_{y}\otimes\sigma_{y})\rho_{p}^{*}(\sigma_{y}\otimes\sigma_{y})$.
For the density matrix in Eq.(\ref{ro10}) the concurrence is given by
\begin{equation}
C(\rho_{p})=\max\{0,2(|d|-\sqrt{p_{00}p_{11}}),2(|d'|-\sqrt{p_{10}p_{01}})\}.
\end{equation}
Fig. 3(a) shows the concurrence as a function of mean photon number $n=\frac{n_0+n_1}{2}$ after squeezing for different values of attenuation $\eta$. One notes the high sensitivity of the concurrence to the attenuation. However, in practice it should be possible to keep losses between the two non-linear crystals very low, values of $\eta$ as high as 0.99 should be realistic. One promising approach would be to realize amplification and de-amplification in a single solid-state system, where the two active non-linear sections could be separated by a non-active spacer layer. Note that the experiment would typically be performed with femtosecond pulses, so it would not be difficult to make non-active layer thick enough to contain the entire pulse. The macro component of the micro-macro entangled state would then exist for a short span of time in that spacer layer. Fig. 3(b) shows that the success probability for projecting the system into the zero-or-one photon subspace decreases as the amount of loss and the mean photon number are increased, but that it is still quite significant in the regime under consideration.

\begin{figure}
\epsfig{file=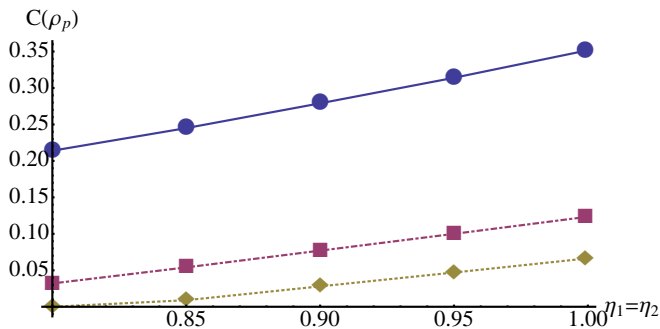,width=1\columnwidth}
\caption{\label{fig4}   Concurrence of the final state for a fixed mean photon number $n=100$ for different values of $\eta$, for $\eta_{1}=\eta_{2}$ varying together.  Circles, cubes and diamonds correspond to $\eta=0.99, 0.97,0.95$ respectively. One can see that the entanglement is much more sensitive to $\eta$ than to $\eta_1$ and $\eta_2$.}
\end{figure}

Fig. 3 suggests that micro-macro entanglement involving hundreds of photons might be observable with the proposed scheme. To confirm this suggestion, it is still important to study the effect of losses before the amplification and after the de-amplification, $\eta_1$ and $\eta_2$. These losses are harder to minimize in practice. In particular, $\eta_2$ also includes detector inefficiency. However, values of order $\eta_1=\eta_2=0.9$ should be achievable \cite{schnabel}. Figure 4 shows that a substantial amount of entanglement is still present in the system under these conditions for a mean photon number $n=100$. One can see that the micro-macro photon-number entanglement is much less sensitive to $\eta_{1}$ and $\eta_{2}$ than to $\eta$, similarly to the results of Ref. \cite{Raeisi12} for micro-macro polarization entanglement.

The entanglement can be demonstrated experimentally by using full homodyne tomography, which was already used to demonstrate the non-locality of single-photon entanglement in Ref. \cite{Lvovsky04}. In this method the full density matrix is reconstructed from the joint quadrature statistics $p_{\theta_{A},\theta_{B}}(X_{A},X_{B})$ for different values of the local oscillator phases $\theta_{A}$ and $\theta_{B}$. In particular this allows one to reconstruct the density matrix in the zero and one-photon subspace which is relevant for us here.

We have proposed to create and detect micro-macro photon-number entanglement by amplifying and then de-amplifying a single-photon entangled state. In particular, the present approach should allow the creation and detection of entangled states that are superpositions of two components with very different mean photon numbers (for example, 50 and 150).

{\it Acknowledgments.} We thank S.Raeisi, S.Rahimi Keshari and B. Sanders  for useful discussions. This work was supported by AITF and NSERC.

\end{document}